\documentclass[11pt]{article}
\usepackage{amsmath,amssymb,amsthm,amsxtra,overpic,bbm,bm,epsfig,subfigure}
\usepackage{color}
\usepackage{comment}

\def\thefootnote{\fnsymbol{footnote}}

\begin{document}

\vspace{0.2cm}

\begin{center}
{\Large\bf Determination of neutrino mass ordering in future $^{\bf 76}${\bf Ge}-based neutrinoless double-beta decay experiments}
\end{center}

\vspace{0.2cm}

\begin{center}
{\bf Jue Zhang $^{a}$} \footnote{E-mail: zhangjue@ihep.ac.cn}
\quad {\bf Shun Zhou $^{a,~b}$} \footnote{E-mail: zhoush@ihep.ac.cn}
\\
{$^a$Institute of High Energy Physics, Chinese Academy of
Sciences, Beijing 100049, China \\
$^b$Center for High Energy Physics, Peking University, Beijing 100080, China}
\end{center}

\vspace{1.5cm}

\begin{abstract}
Motivated by recent intensive experimental efforts on searching for neutrinoless double-beta decays, we perform a detailed analysis of the physics potential of the experiments based on $^{76}\mathrm{Ge}$. Assuming no signals, current and future experiments could place a $90\%$ lower limit on the half life $T^{0\nu}_{1/2} \gtrsim 4\times 10^{26}~{\rm yr}$ and $T^{0\nu}_{1/2} \gtrsim 7\times 10^{27}~{\rm yr}$, respectively. Then, how to report an evidence for neutrinoless double-beta decays is addressed by following the Bayesian statistical approach. For the first time, we present a quantitative description of experimental power to distinguish between normal and inverted neutrino mass orderings. Taking an exposure of $10^{4}~{\rm kg}\cdot{\rm yr}$ and a background rate of $10^{-4}~{\rm counts}/({\rm keV}\cdot{\rm kg}\cdot{\rm yr})$, we find that a moderate evidence for normal neutrino mass ordering (i.e., with a Bayes factor ${\cal B}$ given by $\ln({\cal B}) \simeq 2.5$ or a probability about $92.3\%$ according to the Jeffreys scale) can be achieved if the true value of effective neutrino mass $m^{}_{\beta\beta}$ turns out to be below $0.01~{\rm eV}$.
\end{abstract}

\begin{flushleft}
\hspace{0.8cm} PACS number(s): 11.30.Fs, 12.15.Ff, 23.40.Bw
\end{flushleft}

\def\thefootnote{\arabic{footnote}}
\setcounter{footnote}{0}

\newpage

\section{Introduction}

Tremendous experimental efforts have been made to search for neutrinoless double-beta decays ($0\nu \beta \beta$) in the past decades~\cite{Furry:1939qr,Barabash:2011mf}. The $0 \nu \beta \beta$ process is of particular importance in particle physics, since it is currently the unique feasible way to establish whether neutrinos are their own antiparticles, namely, Majorana particles~\cite{Majorana:1937vz}. The Majorana nature of neutrinos clearly indicates the violation of lepton number, which is accidentally conserved in the standard model of elementary particles (SM). Furthermore, one can also extract the information on the absolute scale of neutrino masses and constrain the Majorana-type CP-violating phases from the observation of $0 \nu \beta \beta$, which can not be obtained from neutrino oscillation experiments~\cite{Agashe:2014kda}. Hence, the establishment of Majorana nature of neutrinos, the determination of absolute neutrino masses, and the constraints on Majorana CP-violating phases will greatly help us explore the origin of neutrino masses and flavor mixing, and pave the way to new physics beyond the SM~\cite{Bilenky:2014uka,Pas:2015eia}.

If neutrinos are indeed Majorana particles, it will be possible to measure a half life $T_{1/2}^{0\nu}$ of some nuclear isotope in the $0\nu\beta\beta$ mode, i.e., $N(A,Z) \to N(A, Z+2) + 2e^-$ with $A$ and $Z$ being mass number and atomic number of the isotope, respectively. The half life is related to the effective neutrino mass $m_{\beta\beta}^{}$ via
\begin{eqnarray}\label{eq:rate}
\left ( T_{1/2}^{0\nu} \right)^{-1} = G_{0\nu}^{} ~|\mathcal{M}_{0\nu}^{}|^2 ~m_{\beta\beta}^2 \; ,
%     (1)
\end{eqnarray}
where $G^{}_{0\nu}$ stands for the phase-space factor \cite{Kotila:2012zza}, and $\mathcal{M}_{0\nu}^{}$ for the relevant nuclear matrix element (NME). Both the phase-space factor and NME can be calculated theoretically, although currently the latter still suffers large uncertainties from nuclear models~\cite{Engel:2015wha}. In addition, the effective neutrino mass is defined as $m^{}_{\beta \beta} \equiv |U^2_{e1} m^{}_1 + U^2_{e2} m^{}_2 + U^2_{e3} m^{}_3|$, where $U^{}_{e i}$ (for $i = 1, 2, 3$) denote the matrix elements in the first row of lepton flavor mixing matrix $U$, and $m^{}_i$ (for $i = 1, 2, 3$) are neutrino masses. For Majorana neutrinos, the $3\times 3$ unitary matrix $U$ is usually parametrized in terms of three mixing angles $\{\theta^{}_{12}, \theta^{}_{13}, \theta^{}_{23}\}$, one Dirac CP phase $\delta$, and two Majorana CP phases $\{\varphi^{}_1, \varphi^{}_2\}$. Taking the standard parametrization of $U$~\cite{Agashe:2014kda} and properly redefining the charged-lepton fields, we obtain
\begin{eqnarray} \label{eq:mbb}
m_{\beta\beta}^{} = |m_1^{}\cos^2\theta_{12}\cos^2\theta_{13} e^{2\mathrm{i} \varphi_1^{}} + m_2^{}\sin^2\theta_{12}\cos^2\theta_{13}  e^{2\mathrm{i} \varphi_2^{}} + m_3^{} \sin^2\theta_{13}| \; ,
%     (2)
\end{eqnarray}
where the mixing angle $\theta^{}_{23}$ and the Dirac CP-violating phase $\delta$ turn out to be irrelevant. In Eq.~(\ref{eq:mbb}), it has actually been assumed that the exchange of light Majorana neutrinos is the dominant mechanism for the $0 \nu \beta \beta$ process to take place, and any other new-physics contributions are negligible. The effective mass $m_{\beta\beta}^{}$ connects the half life $T_{1/2}^{0\nu}$ to the fundamental physical parameters, namely, neutrino mixing angles, neutrino masses, and Majorana CP-violating phases. As an interesting consequence of such a connection, one can determine the currently unresolved neutrino mass ordering by a precise measurement of $T_{1/2}^{0\nu}$. It should be emphasized that this discrimination of mass ordering is only possible under the assumption that neutrinos are Majorana particles. If neutrinos are of Dirac nature, no signals would be observed in $0\nu \beta\beta$ experiments and thus no information on mass ordering can be extracted. Based on current knowledge of neutrino mixing angles and mass-squared differences~\cite{Agashe:2014kda}, one can immediately find that if neutrinos have an inverted mass ordering (IO), i.e., $m^{}_3 < m^{}_1 < m^{}_2$, $m_{\beta\beta}^{}$ has a lower limit ${m^{}_{\beta \beta}}|^{\rm IO}_{\rm min} \approx 0.015~\text{eV}$. No lower bound exists for normal neutrino mass ordering (NO), i.e., $m_1^{} < m_2^{} < m_3^{}$, and the minimum of $m^{}_{\beta \beta}$ could be vanishingly small due to the intricate cancellation among three terms on the right-hand side of Eq.~(\ref{eq:mbb}). See, e.g., Ref.~\cite{Xing:2015zha}, for a detailed analysis. If the cancellation indeed happens, it is impossible to determine the Majorana nature of neutrinos even in the far-future $0\nu\beta\beta$ experiments.

As the determination of neutrino mass ordering is fundamentally important and very challenging for neutrino oscillation experiments, it is intriguing if the $0 \nu \beta \beta$ experiments could provide valuable information. At present, no observation of $0 \nu \beta \beta$ leads to a lower bound on the half life $T^{0\nu}_{1/2}$, which can be translated into an upper bound ${m_{\beta \beta}^{}}|^{\rm exp}$ on the effective neutrino mass, namely, $m^{}_{\beta \beta} < {m_{\beta \beta}^{}}|^{\rm exp}$. Naively speaking, one may claim that the IO is excluded if ${m_{\beta \beta}^{}}|^{\rm exp} < {m^{}_{\beta \beta}}|^{\rm IO}_{\rm min}$ is reached in future. Obviously, it is necessary to put such a claim on a solid statistical ground, and clarify how current and future $0 \nu \beta \beta$ experiments can distinguish IO from NO. This is the primary motivation for the present work.

There have been enormous experimental efforts spent on searching for $0\nu\beta\beta$, using various nuclear isotopes, such as $^{76}\mathrm{Ge}$, $^{136}\mathrm{Xe}$ and $^{130}\mathrm{Te}$. See, e.g., Refs.~\cite{Bilenky:2014uka,Pas:2015eia,GomezCadenas:2011it,Vergados:2012xy,Maneschg:2015dja} for recent reviews on this subject. In this paper, however, we restrict ourselves to the $^{76}\mathrm{Ge}$-based experiments. There are two main reasons for this choice. First, since the $^{76}\mathrm{Ge}$-based experiment possesses an excellent energy resolution, we have a fairly simple background, which has been found to be nearly flat in the signal region. Hence it is rather straightforward to perform projections on physics prospects of this kind of experiments. Second, our analysis will serve as a typical example for the other types of $0\nu\beta\beta$ experiments. Once their dominant backgrounds are well known, one can carry out a similar analysis of them as well.\footnote{One may ignore the issue of background modelling by performing a ``single-bin'' rate analysis, as was done in \cite{GomezCadenas:2010gs}.}

Currently, two $^{76}$Ge-based $0\nu\beta\beta$ experiments, G{\scriptsize ERDA} and M{\scriptsize AJORANA} D{\scriptsize EMONSTRATOR} \cite{Abgrall:2013rze}, are commissioning. In the first phase of G{\scriptsize ERDA}, a lower limit of $T_{1/2}^{0\nu} > 2.1\times 10^{25}$ yr at $90\%$ confidence level was reported for an exposure of 21.6 kg~$\cdot$~yr and a background index (BI) about $10^{-2}$~\cite{Agostini:2013mzu}, corresponding to a background rate of $10^{-2}$ counts/(keV~$\cdot$~kg~$\cdot$~yr). The G{\scriptsize ERDA} Phase-II and M{\scriptsize AJORANA} D{\scriptsize EMONSTRATOR} are expected to increase the exposure to about 200 kg~$\cdot$~yr each, and to reduce the background down to the $10^{-3}$ level at the same time. Such an improvement would increase the current lower limit on $T_{1/2}^{0\nu}$ by one order of magnitude, as shown in Ref.~\cite{Agostini:2015dna}. More excitingly, these two collaborations are also discussing the possibility of building a future large scale $^{76}$Ge (LSGe) experiment together \cite{LSGe}, which may eventually reach an exposure of around $10^4$ kg~$\cdot$~yr, and a reduction of the background index by another order of magnitude, namely, ${\rm BI} = 10^{-4}$. Stimulated by these experimental activities, we will study the physics prospects for the ongoing and upcoming $^{76}{\rm Ge}$-based $0\nu\beta\beta$ experiments. In a recent work \cite{Agostini:2015dna}, an analysis of the projected exclusion limit has been performed. Now we extend that work in two aspects. First, we include a detailed discussion on the experimental potential of discovering a non-null signal. Second, given the possibility of building a tonne-scale experiment LSGe in the future, we make the first attempt to find out at which level the currently undetermined neutrino mass ordering can be resolved then.

The remaining part is structured as follows. To establish our notations and specify the framework of our discussions, we present some general remarks in Section 2, including a brief description of $^{76}{\rm Ge}$-based $0\nu\beta\beta$ experiments, the implemented method for numerical simulations, and an introduction to the Bayesian analysis. In Section 3, we apply the Bayesian method to set up the exclusion limits, discover a non-null signal and discriminate neutrino mass orderings in current and future experiments, following the approach in Ref.~\cite{Caldwell:2006yj}. Finally we summarize our main results in Section 4.

\section{General Remarks}

\subsection{Description of $^{\bf 76}${\bf Ge}-based experiments}

In a real $0\nu\beta\beta$ experiment, we determine the half life $T_{1/2}^{0\nu}$ by collecting events in a narrow energy window, where the signals of our interest should appear. Given a total exposure $\mathcal{E}$ (i.e., a product of the fiducial detector mass and the live time) and a detection efficiency $\epsilon$, the expected number of signal events reads
\begin{eqnarray} \label{eq:Ge}
N^{0\nu}_{} = \ln 2 \cdot N_{\rm A}^{} \cdot\frac{ \mathcal{E} \cdot \epsilon}{M_{\mathrm{Ge}}^{} \cdot T_{1/2}^{0\nu}} \; ,
\end{eqnarray}
with $N_{\rm A}^{} = 6.022\times 10^{23}$ being the Avogadro's constant, and $M_{\mathrm{Ge}} = 75.6$ g the molar mass of $^{76}$Ge. Ideally, all the $0\nu\beta\beta$ events can be identified by the total energy of two final-state electrons, which should show up at the single energy scale 2039 keV, i.e., the $Q$-value of $0\nu\beta\beta$ for $^{76}$Ge. However, because of a finite energy resolution, usually denoted by FWHM (Full Width Half Maximum), in realistic experiments, the observed signal events have a Gaussian distribution, with the mean at the $Q$-value and the width determined by FWHM. The $^{76}$Ge-based $0\nu\beta\beta$ experiments have outstanding energy resolutions and high detection efficiencies. Although the exact values of these parameters vary for different experiments, they are close to each other. For simplicity, we choose FWHM = 3 keV and $\epsilon = 0.65$ as a benchmark point.
%%%%%%%%%%%%%%%%%%%%%%%%%%%%%%%%%%%%% Fig. 1 %%%%%%%%%%%%%%%%%%%%%%%%%%%%%%%
\begin{figure}
\begin{center}
\includegraphics[scale=0.75]{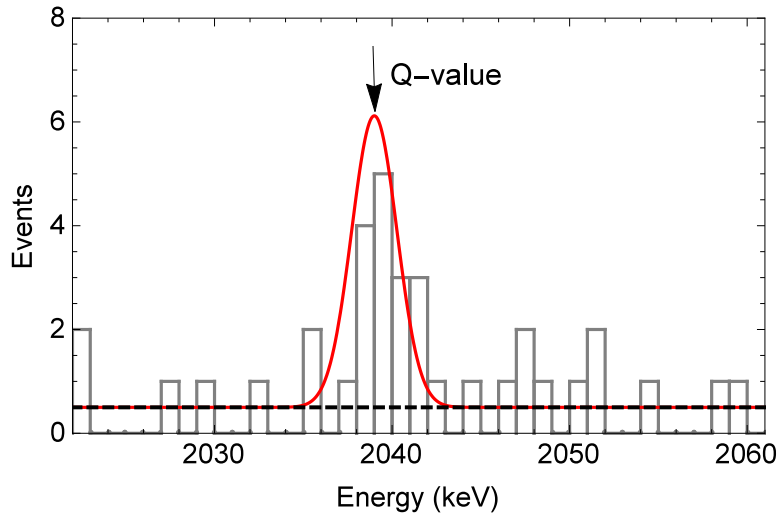}
\end{center}
\vspace{-0.5cm}
\caption{A simulated event spectrum of $0 \nu \beta \beta$ for the $^{76}{\rm Ge}$-based experiment, where the red solid curve represents the expected Gaussian distribution of signal events on top of a constant background and typical values of the half life $T^{0\nu}_{1/2} = 10^{25}~{\rm yr}$, the exposure ${\cal E} = 50~{\rm kg} \cdot {\rm yr}$, the energy resolution ${\rm FWHM} = 3~{\rm keV}$, and the efficiency $\epsilon = 0.65$ have been used. The black-dashed horizontal line corresponds to the background with ${\rm BI} = 10^{-2}$. The gray histograms stand for the total number of both signal and background events, which is randomly generated according to the Poissonian distribution in each energy bin. }
\label{fg:spectrum}
\end{figure}
%%%%%%%%%%%%%%%%%%%%%%%%%%%%%%%%%%%%%%%%%%%%%%%%%%%%%%%%%%%%%%%%%%%%%%%

Besides the signal events, backgrounds are also present in the signal region. Since the background rate varies among current and future experiments, we categorize them into two classes: the first one has a BI of $10^{-3}$ for current experiments, while $10^{-4}$ for the future LSGe experiment. In addition, we assume a flat background model in the energy window of interest, which is quite reasonable because of an excellent energy resolution. Regarding a realistic choice of the exposure, we follow Ref.~\cite{Agostini:2015dna} and take ${\cal E}$ to be up to 400 kg~$\cdot$~yr for current experiments in total, while a maximum of $10^4$ kg~$\cdot$~yr for the future one.

In Fig.~\ref{fg:spectrum}, we show an example of the event spectrum at a $^{76}$Ge-based experiment. The red solid curve represents the Gaussian signal events on top of a constant background (i.e., the black dashed line). Additionally, the gray histograms stand for the total events in each energy bin, which are randomly generated according to the Poissonian distribution. The data have been simulated by assuming a half life of $T^{0\nu}_{1/2} = 10^{25}~{\rm yr}$, ${\rm BI} = 10^{-2}$, and ${\cal E} = 50~{\rm kg}\cdot{\rm yr}$. The benchmark values of the energy resolution and detection efficiency have been used, and the width of energy bins is $1~{\rm keV}$. The simulated number of events, including both signals and backgrounds, is to be fitted by the model parameter (i.e., the half life $T^{0\nu}_{1/2}$ or equivalently the effective neutrino mass $m^{}_{\beta \beta}$), and the details of numerical simulations and statistical analysis will be given in the following two subsections.

\subsection{Numerical simulations}

As shown in Fig.~\ref{fg:spectrum}, we will generate a set of pseudo-data and perform projections with them. Following G{\scriptsize ERDA}'s analysis on its first phase data \cite{Agostini:2013mzu}, we consider a region of spectrum that spans from 2022 keV to 2061 keV with a bin size of 1 keV. The samples of pseudo-data are generated in two different ways. In the first one we simply let the number of events in each energy bin be the expected value from the background and signal functions. The data set produced in this way is known as the Asimov data~\cite{Cowan:2010js}. Once the data are available, one can carry out a statistical analysis to test different hypotheses, or to estimate model parameters by fitting the data.

We also produce the pseudo-data samples by using the Monte Carlo method. More explicitly, the events in each energy bin are randomly generated by assuming a Poissonian distribution with the same expected number of events as for the Asimov data. In this case, a large number of samples are first generated, and each of them is subsequently implemented to test hypotheses or extract model parameters. As a consequence, the statistical distributions of desired quantities can be obtained. This approach is referred to as an ensemble test.

Apparently, no statistical fluctuations are considered in the Asimov data set. However, it has been found in Ref.~\cite{Cowan:2010js} that the analysis of the Asimov data yields a good approximation to the \emph{median} projection of experiments. Therefore, one can employ it to get some quick and qualitative results without the burden of analyzing lots of Monte Carlo samples, as it would be in the case of an ensemble test. For this reason, in the following discussions, we first perform a qualitative projection with the Asimov data set, and support the results by an ensemble test. Next, we present the detailed numerical results with the help of the M{\scriptsize ULTI}N{\scriptsize EST} program \cite{Feroz:2008xx,Feroz:2007kg,Feroz:2013hea}. The sampling is efficiently carried out by using M{\scriptsize ULTI}N{\scriptsize EST}, so is the Bayesian analysis of simulated data.

\subsection{Bayesian analysis}

Now we give a brief introduction to Bayesian statistical analysis. See, e.g., Refs.~\cite{SSbook,Cowan:1998ji}, for more details. Bayesian analysis resides on the well-known Bayes' theorem, and describes the degree of belief in a certain hypothesis $\mathcal{H}^{}_i$ (for $i = 1, ..., r$), given the data set $\mathcal{D}$. Note that the hypotheses ${\cal H}^{}_i$ are mutually exclusive, and only one of them is actually true. According to the Bayes' theorem, the posterior probability of the hypothesis $\mathcal{H}^{}_i$, denoted as $\mathrm{Pr}(\mathcal{H}^{}_i|\mathcal{D})$, is given by
\begin{eqnarray} \label{eq:bayes}
\mathrm{Pr}(\mathcal{H}^{}_i|\mathcal{D}) = \frac{\mathrm{Pr}(\mathcal{D}|\mathcal{H}^{}_i) ~\mathrm{Pr}(\mathcal{H}^{}_i)}{\mathrm{Pr}(\mathcal{D})} \; .
\end{eqnarray}
In the above equation, $\mathrm{Pr}(\mathcal{H}^{}_i)$ stands for the prior probability of the hypothesis $\mathcal{H}^{}_i$, reflecting our prior degree of belief in such a hypothesis. $\mathrm{Pr}(\mathcal{D}|\mathcal{H}_i)$ is the probability of obtaining the data $\mathcal{D}$, assuming the hypothesis $\mathcal{H}_i$ to be true, and is called the evidence $\mathcal{Z}_i$ of the hypothesis $\mathcal{H}_i$. Lastly, the overall probability of observing the data $\mathcal{D}$ is denoted by $\mathrm{Pr}(\mathcal{D})$, and it is equal to $\sum_{i=1}^{r} \mathrm{Pr}(\mathcal{D}|\mathcal{H}_i) \mathrm{Pr}(\mathcal{H}_i)$, because of the normalization condition $\sum_{i=1}^{r} \mathrm{Pr}(\mathcal{H}_i|{\cal D}) = 1$.

One direct application of the above formalism is to make model selection. With the help of Eq.~(\ref{eq:bayes}), one can compute the posterior odds of two competing hypotheses by taking the ratio of their posterior probabilities, namely,
\begin{eqnarray}
\frac{\mathrm{Pr}(\mathcal{H}_i|\mathcal{D})}{\mathrm{Pr}(\mathcal{H}_j|\mathcal{D})} = \frac{\mathcal{Z}_i}{\mathcal{Z}_j} \frac{\mathrm{Pr}(\mathcal{H}_i)}{\mathrm{Pr}(\mathcal{H}_j)} \; ,
\end{eqnarray}
where the ratio of evidences ${\cal B} \equiv \mathcal{Z}_i/\mathcal{Z}_j$ is termed Bayes factor. In the case of no prior preference for any hypothesis (i.e., equal prior probabilities), the posterior odds is then directly reflected by the Bayes factor. Moreover, to interpret the value of this posterior odds or the Bayes factor, one often adopts the Kass-Raftery~\cite{KR} or Jeffreys~\cite{Jeffreys,Jeffreys_mod,Trotta:2008qt} scale. In Table~\ref{tb:Jeffreys} we list the Jeffreys scale that used in \cite{Jeffreys_mod,Trotta:2008qt}, and will implement them to interpret the results in the present work.
\begin{table}
\centering
\begin{tabular}{l | l | l | l}
\hline
\hline
$\left|\ln(\text{odds})\right|$ & Odds & Probability & Interpretation \\
\hline
$< 1.0$ & $\lesssim 3 : 1$ & $\lesssim 75.0\%$ & Inconclusive \\
$1.0$ & $\simeq 3 : 1$ & $\simeq 75.0\%$ & Weak evidence\\
$2.5$ & $\simeq 12 : 1$ & $\simeq 92.3\%$ & Moderate evidence \\
$5.0$ & $\simeq 150 : 1$ & $\simeq 99.3\%$ & Strong evidence\\
\hline
\hline
\end{tabular}
\vspace{0.3cm}
\caption{The Jeffreys scale used for the statistical interpretation of Bayes factors, posterior odds and model probabilities~\cite{Jeffreys_mod, Trotta:2008qt}.}
\label{tb:Jeffreys}
\end{table}

Another realm of applying the Bayesian analysis is the parameter estimation. Now suppose that there exists a single hypothesis $\mathcal{H}$ that describes the data well, and this hypothesis is specified by a set of free parameters $\Theta$. The posterior probability distribution of $\Theta$ is then calculated as
\begin{eqnarray}
\mathrm{Pr}(\Theta|\mathcal{D, H}) = \frac{\mathrm{Pr}(\mathcal{D}|\Theta, \mathcal{H}) ~\mathrm{Pr}(\Theta|\mathcal{H})}{\mathrm{Pr}(\mathcal{D}|\mathcal{H})} = \frac{\mathcal{L}(\Theta)~\pi(\Theta)}{\mathcal{Z}} \; ,
\end{eqnarray}
where the likelihood function $\mathrm{Pr}(\mathcal{D}|\Theta, \mathcal{H})$ and the prior probability distribution $\mathrm{Pr}(\Theta|\mathcal{H})$ are usually also denoted as $\mathcal{L}(\Theta)$ and $\pi(\Theta)$, respectively. Similarly, because of the normalization condition for the posterior probability, the evidence $\mathcal{Z}$ is found to be
\begin{eqnarray}
\mathcal{Z} = \int \mathrm{Pr}(\mathcal{D}|\Theta, \mathcal{H}) ~\mathrm{Pr}(\Theta|\mathcal{H})~ \mathrm{d}^N\Theta = \int \mathcal{L}(\Theta)~\pi(\Theta) ~ \mathrm{d}^N\Theta \; ,
\end{eqnarray}
where $N$ is the dimension of parameter space. Note that in the parameter estimation, the evidence $\mathcal{Z}$ is a common factor for all parameters in $\Theta$, and thus can be ignored. This is not the case if some other hypotheses are also considered, as we have seen from the discussions of model selection. The posterior distribution of $\Theta$ then essentially depends on two factors, the likelihood function $\mathcal{L}(\Theta)$ and the prior distribution $\pi(\Theta)$, and one may view this parameter estimation as an update on the prior distribution in light of new data. With the posterior distribution of the parameter, one can extract the most probable value and the credible regions in the standard way~\cite{SSbook}.

We now discuss the application of the above general formalism to the $0\nu \beta \beta$ experiments. For concreteness, let us consider the analysis of setting up the exclusion limit, which belongs to the category of parameter estimation. The essential part of this analysis is to obtain the posterior distribution of the number of signal events $N^{0\nu}_{}$ by fitting the simulated spectrum as given in Fig.~\ref{fg:spectrum}. In the fit, our parameters of interest $\Theta$ are the number of signal events $N^{0\nu}_{}$ and the possible background events $B$, i.e., $\Theta = \{N^{0\nu}, B\}$. Both $N^{0\nu}$ and $B$ are drawn from some uniform prior distributions $\pi(N^{0\nu}, B)$. With each specified set of $\Theta$, we first calculate the expected number of events in each bin $\lambda_i^{}(N^{0\nu}, B)$, where $i$ is the bin index, by assuming a Gaussian distribution of the signal events and a flat background (exactly in the same way as we generate the simulated spectrum in Fig.~\ref{fg:spectrum}). Then, we compare the expected events $\lambda_i$ with the simulated data events $n_i$ in each bin, and obtain the likelihood $\mathcal{L}(N^{0\nu}, B)$ of this chosen set $\{N^{0\nu}, B\}$ by assuming a Poisson distribution, i.e.,
\begin{eqnarray} \label{eq:likelihood}
\mathcal{L}(N^{0\nu}, B) = \prod_{i=1}^{N^{}_{\rm b}} \frac{\lambda_i^{n_i}}{n_i^{} !} e^{-\lambda_i^{}},
\end{eqnarray}
where $N^{}_{\rm b}$ is the total number of bins. With the prior distribution $\pi(N^{0\nu}_{}, B)$, we find the two-dimensional distribution ${\rm Pr}(\{N^{0\nu}_{}, B\} | \mathcal{D}) =  \mathcal{L}(N^{0\nu}, B)~\pi(N^{0\nu}_{}, B)/{\cal Z}$, from which the posterior distribution of $N^{0\nu}$ can be obtained by marginalizing over the background events, namely, ${\rm Pr}(N^{0\nu}_{} | \mathcal{D}) = \int {\rm Pr}(\{N^{0\nu}_{}, B\} | \mathcal{D}) ~{\rm d}B$. Based on the posterior distribution, one can derive the exclusion limits on $T^{0\nu}_{1/2}$ and $m^{}_{\beta \beta}$ via Eq.~(\ref{eq:Ge}) and Eq.~(\ref{eq:rate}), respectively.

For the claim of discovering the $0\nu\beta \beta$ decays and the discrimination of neutrino mass orderings, the same spirit as above will be followed, except that those two analyses belong to hypothesis test and the fitting parameters are different. For instance, the model parameters of our interest in the discrimination of neutrino mass orderings would be neutrino mixing parameters. In order to fit the data, however, one has to convert each set of neutrino mixing parameters to the expected number of signal events $N^{0\nu}$ via Eqs.~(\ref{eq:rate}), (\ref{eq:mbb}) and (\ref{eq:Ge}). After doing so, we can similarly calculate the likelihood as defined in Eq.~(\ref{eq:likelihood}), and then obtain the posterior distribution for the chosen set of neutrino parameters. As now we are concerned with hypothesis test, in a final step, we integrate the product of the likelihood function and the prior distribution over the whole allowed range of neutrino parameters to obtain the evidence. By comparing the evidences of two competing hypotheses of neutrino mass orderings, we then know which one is more favorable, given a particular data set.

\section{Prospects for $^{\bf 76}${\bf Ge}-based experiments}

\subsection{Setting up the exclusion limit}

First, we consider how to set up the exclusion limits in current and future $^{76}$Ge-based experiments. To do so, we generate the pseudo-data samples without signal contributions, and derive the posterior distribution of the number of signal events $N^{0\nu}$ by fitting the data. A uniform prior on $N^{0\nu}$ is assumed in the fit, and the background level is allowed to float. As already pointed out in Ref.~\cite{Agostini:2015dna}, including systematic uncertainties yields negligible changes to the final results, so we choose to ignore them as well. The 90$\%$ upper limit on the signal events $N^{0\nu}_{0.9}$ can be obtained according to
\begin{eqnarray}
\int_0^{N^{0\nu}_{0.9}} \mathrm{Pr}(N^{0\nu} | \mathcal{D}) ~\mathrm{d} N^{0\nu} = 0.90 \; ,
\end{eqnarray}
where $\mathrm{Pr}(N^{0\nu} | \mathcal{D})$ is the posterior distribution of $N^{0\nu}$. Furthermore, it is straightforward to translate $N^{0\nu}_{0.9}$ into the 90$\%$ lower limit on $T_{1/2}^{0\nu}$ by using Eq.~(\ref{eq:Ge}).

%%%%%%%%%%%%%%%%%%%%%%%%%%%%%%%%% Fig. 2 %%%%%%%%%%%%%%%%%%%%%%%%%%%%%%%%%%
\begin{figure}
\begin{center}
\includegraphics[scale=0.65]{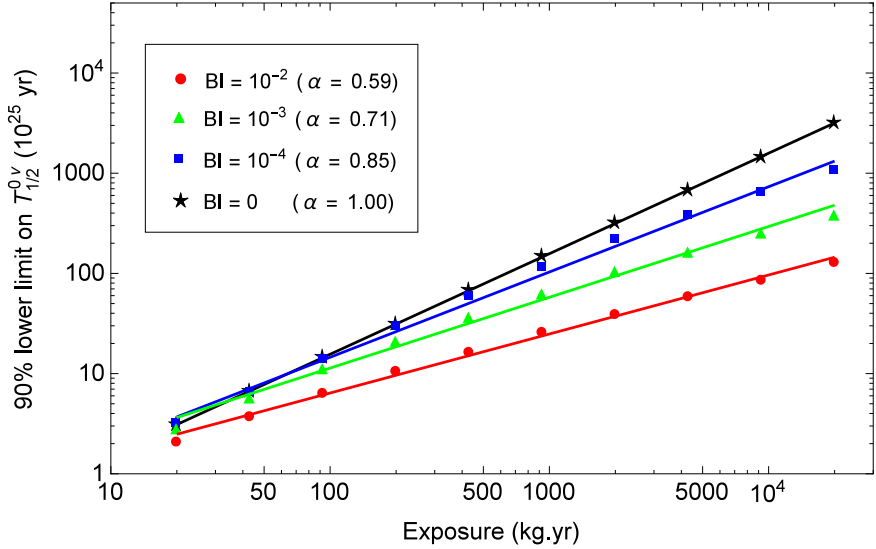}
\end{center}
\vspace{-0.5cm}
\caption{The 90$\%$ lower limit on $T_{1/2}^{0\nu}$ as a function of the total exposure in the case of the Asimov data. Four different background levels are considered, i.e., BI of $10^{-2}$ (red dot), $10^{-3}$ (green triangle), $10^{-4}$ (blue square) and background free (black star). The scalability index $\alpha$ is found by fitting each data set. }
\label{fg:limit_asimov}
\end{figure}
%%%%%%%%%%%%%%%%%%%%%%%%%%%%%%%%%%%%%%%%%%%%%%%%%%%%%%%%%%%%%%%%%%%%%%%%%%
In Fig.~\ref{fg:limit_asimov}, we present the projected $90\%$ lower limit on $T_{1/2}^{0\nu}$ as a function of the total exposure $\mathcal{E}$, where the Asimov data are used. Four different background levels are considered, i.e., BI of $10^{-2}$ (red dot), $10^{-3}$ (green triangle), $10^{-4}$ (blue square) and background free (black star). Some comments are in order:
\begin{itemize}
\item An interesting observation is that for each background level, the projected lower limit $( T_{1/2}^{0\nu} )_{0.9}$ has a good power-law dependency on the exposure, i.e., $( T_{1/2}^{0\nu} )_{0.9} \propto \mathcal{E}^\alpha$, where $\alpha$ is defined as the scalability index (SI). This index characterizes the gain on sensitivity when the experimental setup is expanded to a larger scale. From Fig.~\ref{fg:limit_asimov}, one can see that a larger BI leads to a lower SI. Consequently, scaling two experimental setups by a common factor, one can conclude that the experiment that has a larger background would have less gain on the sensitivity. This demonstrates that both increasing the scale of experiments and reducing the backgrounds are very important to improve the sensitivity.

\item Moreover, one may also notice that the SI is ranging from $0.5$ to $1$ for different background levels. This can be understood in the following way. First of all, let us start with the background-free case. Since the pseudo-data samples are generated without signal contributions, there should be no events at all in all energy bins. For any values of the exposure, one actually fits the same empty data sample, so the upper limit $N_{0.9}^{0\nu}$ remains unchanged. With such a fixed $N_{0.9}^{0\nu}$, we find that the lower limit on $T_{1/2}^{0\nu}$ is indeed proportional to the exposure, according to Eq.~(\ref{eq:Ge}). Therefore, the SI of $\alpha = 1$ in the background-free case is understood.

    Then, we come to nonzero backgrounds. Since a data sample with backgrounds has sensitivities only to the signal that is large enough to manifest itself as the background fluctuation, an upper limit on $N^{0\nu}$ can be obtained when it is on the same order of background fluctuations, which can be characterized by $\sigma_{\mathrm{bkgd}}^{}$, i.e., $N_{0.9}^{0\nu} \sim \sigma_{\mathrm{bkgd}}^{}$. For the value of $\sigma_{\mathrm{bkgd}}^{}$, one may take it to be the standard deviation $\sigma_{\mathrm{bkgd}}^{} = \sqrt{N_\mathrm{bkgd}}$. Since $N_\mathrm{bkgd}$ itself scales linearly with the exposure, we finally obtain $( T_{1/2}^{0\nu} )_{0.9} \propto \sqrt{\mathcal{E}}$. Thus, this explains $\alpha \sim 0.5$ for BI = $10^{-2}$ that is large enough. For the intermediate values of BI, one can interpret the corresponding SI's as they should interpolate between $0.5$ and $1$.
\end{itemize}

\begin{figure}
\begin{center}
\includegraphics[scale=0.58]{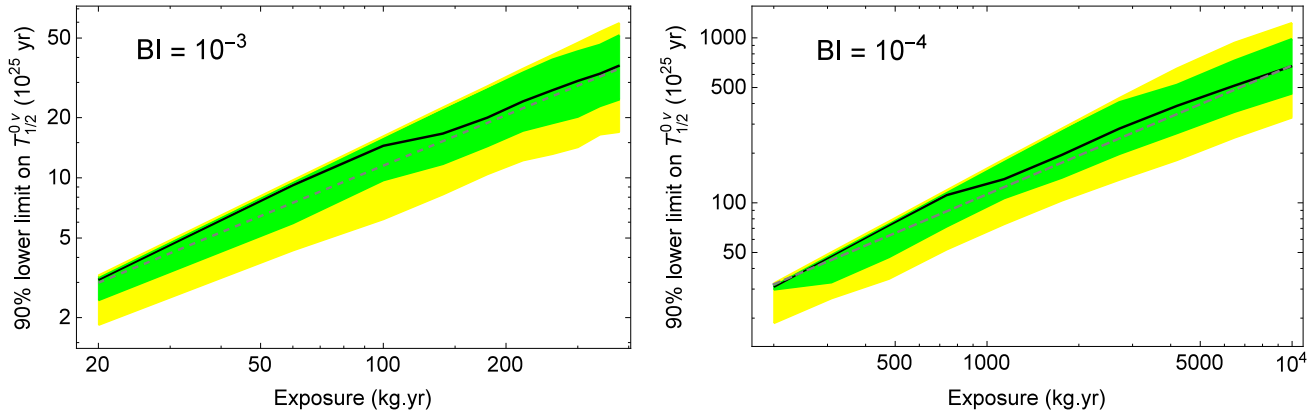}
\end{center}
\vspace{-0.5cm}
\caption{The 90$\%$ lower limit on $T_{1/2}^{0\nu}$ as a function of the total exposure in the case of ensemble test. Left (Right) panel corresponds to the background level of $\mathrm{BI} = 10^{-3}(10^{-4})$. The $1\sigma$ and $2\sigma$ bands are denoted by the green and yellow regions, while the median values are represented by the black curves. For comparison, we also show the results from the Asimov data set by gray dashed lines.}
\label{fg:limit_MC}
\end{figure}

The results from ensemble tests are given in Fig.~\ref{fg:limit_MC}, where we just consider the cases with BI of $10^{-3}$ and $10^{-4}$ that respectively represent the current and future experimental setups, and realistic ranges of the exposure are chosen. For both cases, the median projection on the lower limit of $T_{1/2}^{0\nu}$ (black curve) together with its $1\sigma$ (green) and $2\sigma$ (yellow) bands is mapped out. For comparison, we also depict the results from the Asimov data set, corresponding to the gray dashed curve. As one can see, the Asimov data set yields rather good approximation to the median performance. Finally, one finds that for the current generation of experiment, a median exclusion limit of $T^{0\nu}_{1/2} \gtrsim 4\times 10^{26}$ yr can be obtained, while $T^{0\nu}_{1/2} \gtrsim 7\times 10^{27}$ yr for the future one. Our conclusions are rather close to those given in \cite{Agostini:2015dna}.
%%%%%%%%%%%%%%%%%%%%%%%%%%%%%%%%%%%%%%%%%%% Fig. 3 %%%%%%%%%%%%%%%%%%%%%%%%
%%%%%%%%%%%%%%%%%%%%%%%%%%%%%%%%%%%%%%%%%%%%%%%%%%%%%%%%%%%%%%%%%%%%%%%%%%%%

\subsection{Discovering a non-null signal}

Then, we proceed with the analysis of discovering a non-null signal. Unlike the analysis of exclusion limits, we generate samples with the signal contributions, and analyze them so as to find out at which significance level the experiments can discover the $0\nu\beta\beta$ signal. In practice, the signal contributions in data are calculated by inputting different true values of $T_{1/2}^{0\nu}$. For each true value of $T_{1/2}^{0\nu}$ we perform a Bayesian hypothesis test with the following two competing hypotheses:
\begin{eqnarray}
\mathcal{H} &:& \text{the sample spectrum is due to background only} \; ; \nonumber \\
\overline{\mathcal{H}} &:& \text{the sample spectrum contains both signal and background contributions}\; . \nonumber
\end{eqnarray}
The prior probabilities of $\mathcal{H}$ and $\overline{\mathcal{H}}$ are assumed to be equal, and a moderate (strong) evidence for a signal will be claimed if the logarithm of the Bayes factor $\ln \mathcal{B} = \ln (\mathcal{Z}_{\bar{\mathcal{H}}} / \mathcal{Z}_{\mathcal{H}})$ is larger than 2.5 (5.0), according to the Jeffreys scale in Table \ref{tb:Jeffreys}.

\begin{figure}
\begin{center}
\includegraphics[scale=0.9]{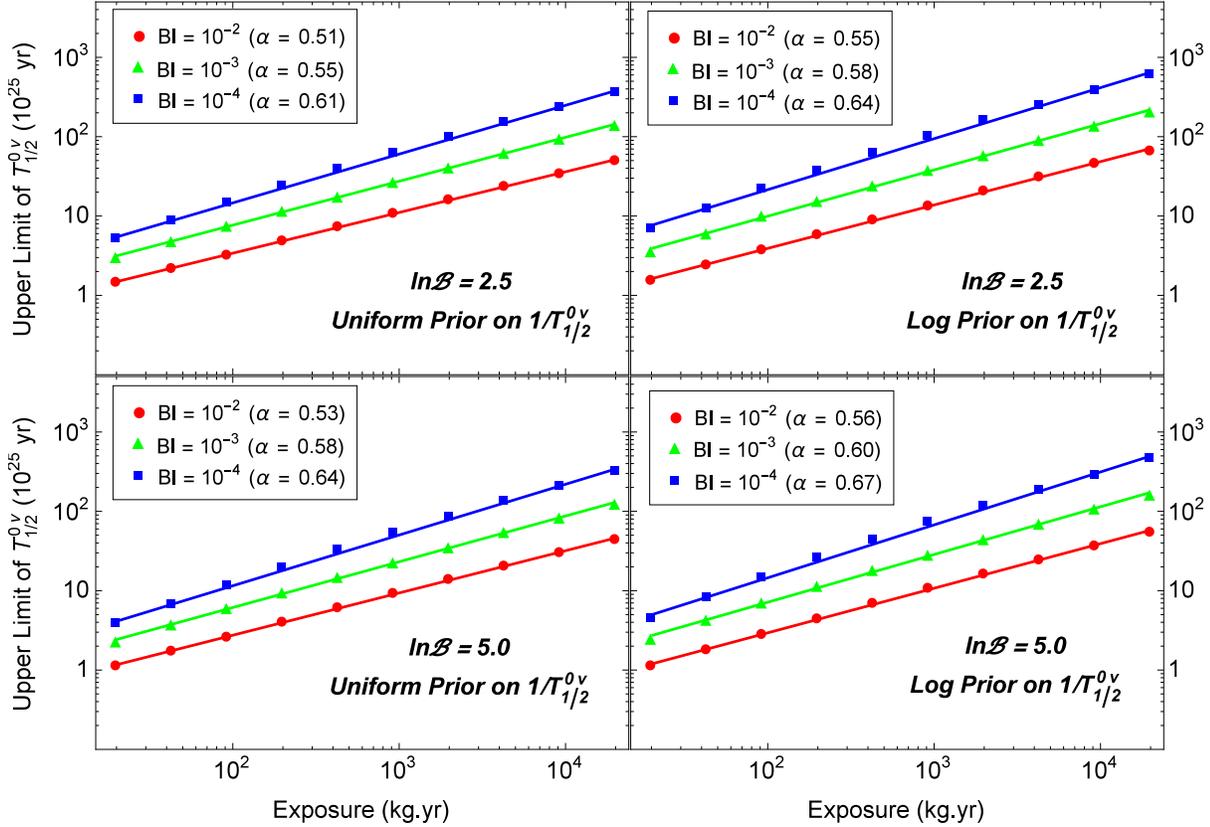}
\end{center}
\vspace{-0.5cm}
\caption{Upper limits on $T_{1/2}^{0\nu}$, below which one has about 50$\%$ probability of reporting a moderate (top panel) or strong (bottom panel) evidence for the $0\nu\beta\beta$ signal, as a function of the exposure. The results have been obtained by using the Asimov data. Three different background levels, i.e., BI of $10^{-2}$ (red dot), $10^{-3}$ (green triangle) and $10^{-4}$ (blue square), are considered. In addition, the left (right) panel corresponds to the case of a uniform (logarithmic) prior on $1/T_{1/2}^{0\nu}$ in the fit.}
\label{fg:discovery_Asimov}
\end{figure}

As before, we first perform the analysis using the Asimov data set. In this case, since no statistical fluctuation is considered, a definite conclusion regarding moderate or strong evidence for a signal can be made for any true value of $T_{1/2}^{0\nu}$. For a given set of exposure and background level, one only needs to present an upper limit on the true value of $T_{1/2}^{0\nu}$, below which one has about 50$\%$ probability of reporting a moderate or strong evidence for the $0\nu\beta\beta$ signal. In Fig.~\ref{fg:discovery_Asimov}, we present the detailed projection on the moderate and strong evidence limits on $T_{1/2}^{0\nu}$ as a function of the exposure. Three different background levels, BI of $10^{-2}$, $10^{-3}$ and $10^{-4}$, are considered. Moreover, in order to investigate the dependence on priors, we adopt two different scenarios, i.e., a uniform prior and a logarithmic prior both for  $1/T_{1/2}^{0\nu} \in [10^{-29}, 10^{-24}]~{\rm yr}^{-1}$.\footnote{Our choice of imposing a prior on $1/T_{1/2}^{0\nu}$ is to follow the analysis done by G{\scriptsize ERDA} \cite{Agostini:2013mzu}. In fact, it is equivalent to doing that on the signal event $N^{0\nu}$ according to Eq.~(\ref{eq:Ge}). Moreover, for the fitting parameter, about which very limited experimental information is currently available, we tend to choose a uniform prior if it is dimensionless, while considering both uniform and logarithmic priors if it has a dimension. Such a treatment is rather conservative and reflects our current ignorance of that parameter. Once the experimental information is available, one may easily incorporate the realistic prior (e.g., a normal distribution) into the analysis.} By carefully inspecting these results, we observe several interesting features. First, the dependence on the priors of $1/T_{1/2}^{0\nu}$ seems to be quite mild, which indicates our results are robust against the choices of priors. Second, for all cases, a good power-law relation between the required maximal value of $T_{1/2}^{0\nu}$ and the exposure is observed, as in the previous case of exclusion limits. Lastly, the limits for moderate and strong evidence cases are quite similar, and only about 20$\%$ discrepancy between them has been observed.
%%%%%%%%%%%%%%%%%%%%%%%%%%%%%%%%%% Fig. 4 %%%%%%%%%%%%%%%%%%%%%%%%%%%%%%%%%%
%%%%%%%%%%%%%%%%%%%%%%%%%%%%%%%%%%%%%%%%%%%%%%%%%%%%%%%%%%%%%%%%%%%%%%%%%%

\begin{figure}
\begin{center}
\includegraphics[scale=0.9]{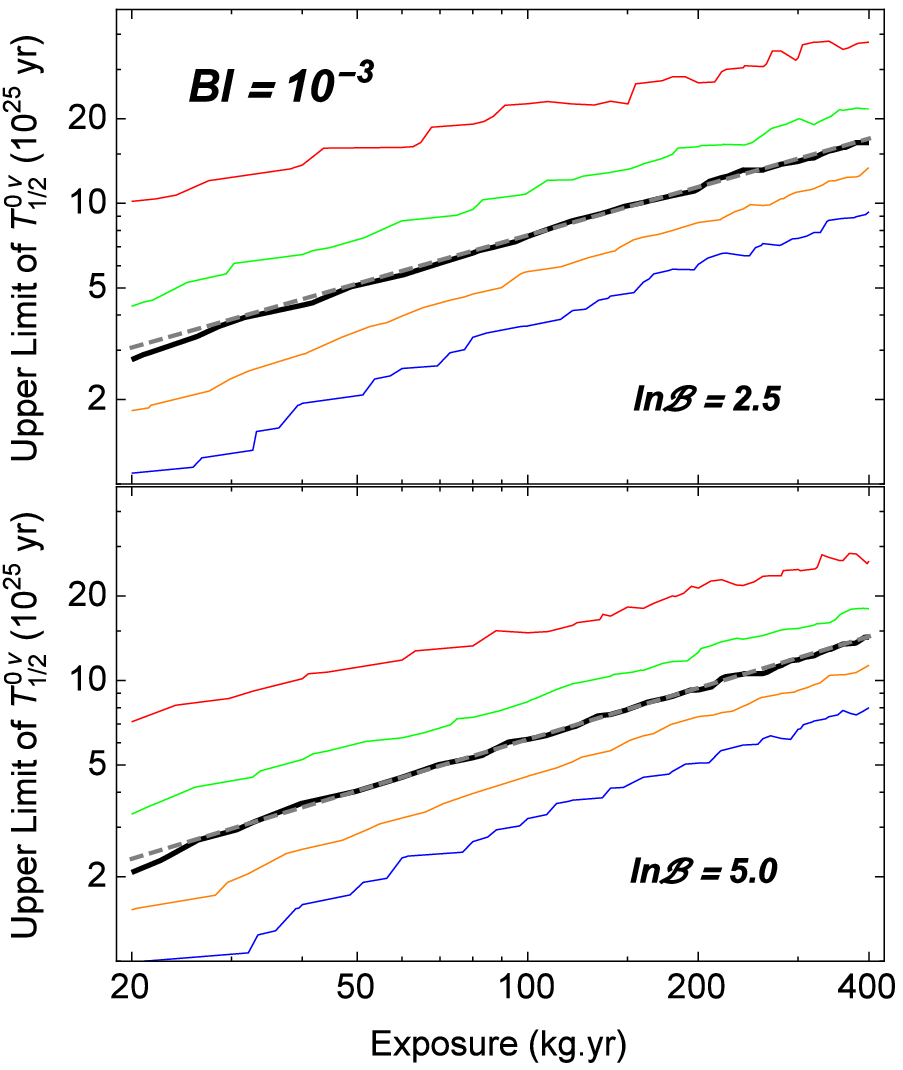}\qquad
\includegraphics[scale=0.92]{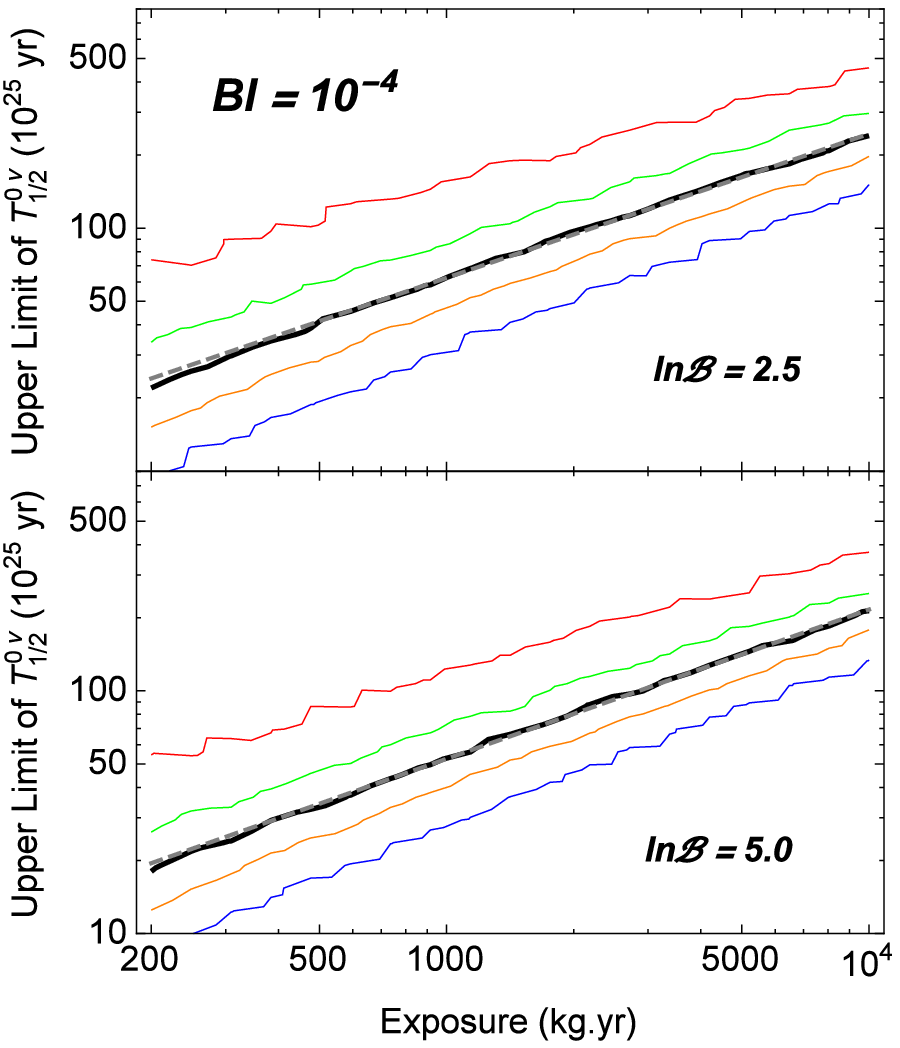}
\end{center}
\vspace{-0.5cm}
\caption{Upper limits on $T_{1/2}^{0\nu}$, at which one may report a moderate (top panel) or strong (bottom panel) evidence for the signal, as a function of the exposure in the case of ensemble test. A uniform prior on $1/T_{1/2}^{0\nu}$ is taken in the fit. Left (Right) panel corresponds to BI of $10^{-3}$ and $10^{-4}$, respectively. For each sub-figure, we show the contours that have probabilities of 5$\%$ (red), 25$\%$ (green), 50$\%$ (black), 75$\%$ (orange) and 95$\%$ (blue) to report a moderate or strong evidence. For comparison, the results from the Asimov data set are given by the gray dashed lines, which are almost indistinguishable from the 50$\%$ black solid curves. }
\label{fg:discovery_MC}
\end{figure}

We next turn to the case of ensemble tests. For each true value of $T_{1/2}^{0\nu}$, one now needs to generate many samples using the Monte Carlo method. Because of statistical fluctuations, we may not reach a consensus about discovery among all samples. Therefore, instead of simply reporting an upper limit on $T_{1/2}^{0\nu}$, we have to associate a probability of reporting a moderate or strong evidence for each $T_{1/2}^{0\nu}$. In Fig.~\ref{fg:discovery_MC}, we present the contours of $T_{1/2}^{0\nu}$ for five different probabilities of claiming a moderate or strong evidence as a function of the exposure. Again, we only consider realistic choices of exposure and background levels. The prior on $1/T_{1/2}^{0\nu}$ is chosen to be uniform, as it has been demonstrated that different priors yield almost identical results.

From Fig.~\ref{fg:discovery_MC}, we observe that the results from the Asimov data (gray dashed curves) agree very well with the contours that have 50$\%$ probability of reporting a moderate or strong evidence. Secondly, the contours for different probabilities also show good power-law relations with respect to the exposure. Their SI's seem to be quite close. Finally, one can report a strong evidence for a half life of $T^{0\nu}_{1/2} \sim 10^{26}$ yr with 50$\%$ probability for current generation of experiments, while $T^{0\nu}_{1/2} \sim 2\times 10^{27}$ yr for the future ones.
%%%%%%%%%%%%%%%%%%%%%%%%%%%%%%%%%%%% Fig. 5 %%%%%%%%%%%%%%%%%%%%%%%%%%%%%%%%
%%%%%%%%%%%%%%%%%%%%%%%%%%%%%%%%%%%%%%%%%%%%%%%%%%%%%%%%%%%%%%%%%%%%%%%%%%%%%

\subsection{Discriminating neutrino mass orderings}

Finally, we are ready to deal with the important issue of discriminating two neutrino mass orderings in the $^{76}$Ge-based $0\nu\beta\beta$ experiments. From previous discussions, we have already seen that current generation of experiments have a sensitivity only to $T_{1/2}^{0\nu} \sim 10^{26}~\mathrm{yr}$, which is too small to tell two mass orderings apart. Therefore, in this subsection, we focus on the future generation of experiments, which has a BI of $10^{-4}$ and an exposure up to $10^4$~kg~$\cdot$~yr.

\begin{figure}
\begin{center}
\includegraphics[scale=0.96]{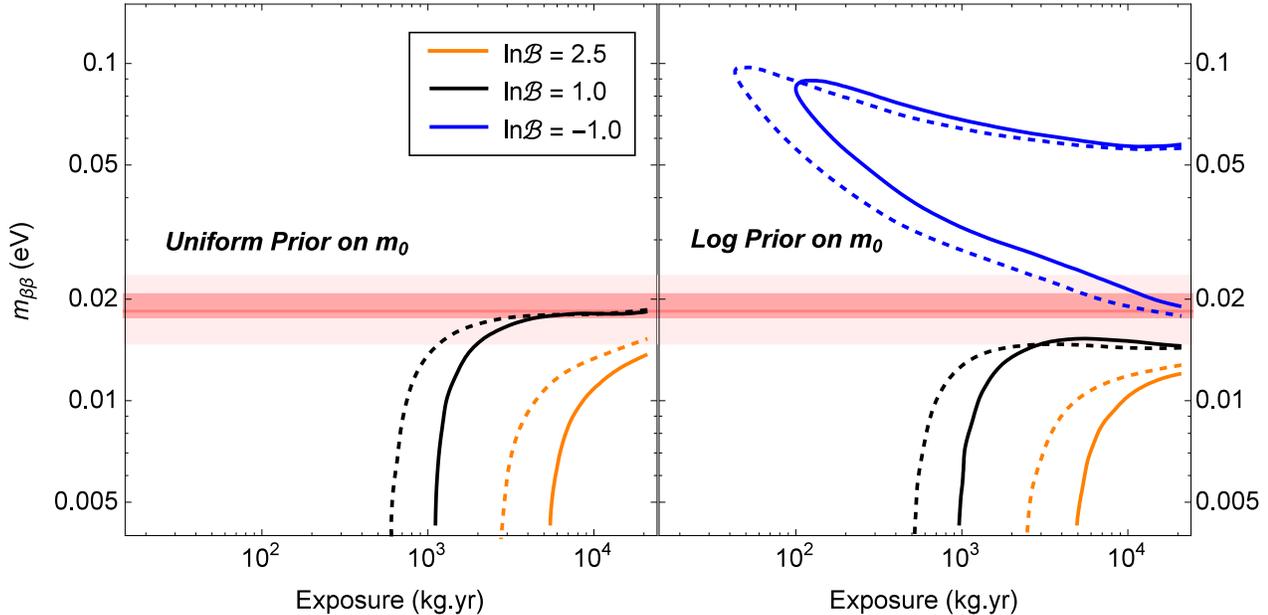}
\end{center}
\vspace{-0.5cm}
\caption{Contours of the Bayes factor $\ln{\cal B} \equiv \ln({\cal Z}^{}_{\rm NO}/{\cal Z}^{}_{\rm IO})$ calculated by comparing between NO with IO, where the thick solid (dashed) curves correspond to the low NME ${\cal M}^{}_{0\nu} = 4.6$ (the high NME ${\cal M}^{}_{0\nu} = 5.8$). The orange, black and blue curves represent $\ln({\cal B}) = 2.5$, $1$ and $-1$, respectively. }
\label{fg:NH_IH}
\end{figure}

In this analysis we are using the Asimov data, which should represent the median case of a full Monte Carlo study. The signal contributions are included by taking various true values of $T_{1/2}^{0\nu}$, as done in the previous discovery analysis. Then, we perform the Bayesian hypothesis test and compare between the evidences of the NO and IO hypotheses. The details of our calculations are summarized as follows:
\begin{enumerate}
\item What we actually need to do is to find out whether NO or IO is preferred, given a simulated data set ${\cal D}$. This can be achieved by calculating the evidences $ \mathcal{Z}_{\rm NO}= \int {\rm Pr}({\cal D}|\Theta, {\rm NO}) \pi (\Theta) \mathrm{d}^{\rm N}_{}\Theta $ and $ \mathcal{Z}_{\rm IO}= \int {\rm Pr}({\cal D}|\Theta, {\rm IO}) \pi (\Theta) \mathrm{d}^{\rm N}_{}\Theta $, where $\Theta$ now stands for neutrino mass and mixing parameters. Therefore, we start with the prior probabilities of neutrino parameters,\footnote{In fact, we do not need $\theta_{23}^{}$ and $\delta$, as the effective neutrino mass $m_{\beta\beta}^{}$ defined in Eq.~(\ref{eq:mbb}) does not depend on them.} i.e., three lepton mixing angles $\theta^{}_{ij}$ (for $ij = 12, 13, 23$), three CP phases $\{\delta, \varphi^{}_1, \varphi^{}_2\}$, two neutrino mass-squared differences and the lightest neutrino mass $m_0^{}$. For the mixing angles and mass-squared differences, which have been precisely measured, we assume Gaussian priors with the central values and $1\sigma$ errors taken from Ref.~\cite{Gonzalez-Garcia:2014bfa}. For CP-violating phases, uniform priors are chosen. Lastly, we adopt two different priors for the lightest neutrino mass $m_0^{}$, i.e., a uniform prior within $[0, 0.2]~\text{eV}$ and a logarithmic prior on $[10^{-5}, 0.2]~\text{eV}$.

\item With neutrino parameters, we are able to calculate the corresponding effective neutrino mass $m_{\beta\beta}^{}$ by using Eq.~(\ref{eq:mbb}). For $^{76}$Ge decays, the phase-space factor $G^{}_{0\nu} = 2.63\times 10^{-25}~{\rm yr}^{-1}~{\rm eV}^{-2}$ has been given in Refs.~\cite{Kotila:2012zza,Guzowski:2015saa}. In order to calculate $T^{0\nu}_{1/2}$ in Eq.~(\ref{eq:rate}) and eventually the number of signal events in Eq.~(\ref{eq:Ge}), we have to specify the values of NME $\mathcal{M}_{0\nu}^{}$, on which currently there exist large uncertainties. For illustration, we consider two limiting cases, namely, $\mathcal{M}_{0\nu}^{} = 4.6$ referring to as the ``Low NME" case while $\mathcal{M}_{0\nu}^{} = 5.8$ as the ``High NME" case.\footnote{Here we follow the same treatment on NME as that given in Ref.~\cite{Agostini:2015dna}, where NME is restricted in the range between 4.6 and 5.8. This range covers most of the calculations available in the literature, except for that in the nuclear shell model, where a rather small value of 2.3 is predicted~\cite{Engel:2015wha}. Adopting the result from the nuclear shell model, we find that the contours of the Bayes factor in Fig.~\ref{fg:NH_IH} would shift further to the right, decreasing the discriminating power on neutrino mass orderings.} Putting all together, we compute the evidences for ${\cal Z}^{}_{\rm NO}$ and ${\cal Z}^{}_{\rm IO}$ in the NO and IO cases, respectively. Since we assume equal priors for these two hypotheses, the ratio of their posterior probability coincides with the ratio of their evidences, i.e., the Bayes factor $\mathcal{B} = {\cal Z}^{}_{\rm NO}/{\cal Z}^{}_{\rm IO}$. Finally, one then uses the Jeffreys scale in Table~\ref{tb:Jeffreys} to interpret such a ratio, and figure out the level of preference for neutrino mass ordering.

\end{enumerate}

%%%%%%%%%%%%%%%%%%%%%%%%%%%%%%%%%%%%% Fig. 6 %%%%%%%%%%%%%%%%%%%%%%%%%%%%%
%%%%%%%%%%%%%%%%%%%%%%%%%%%%%%%%%%%%%%%%%%%%%%%%%%%%%%%%%%%%%%%%%%%%%%%%%

The primary results in this paper have been summarized in Fig.~\ref{fg:NH_IH}, where the contours of Bayes factors in the plane of $m^{}_{\beta \beta}$ and ${\cal E}$ are given. The left and right panels correspond to the uniform and logarithmic priors on the lightest neutrino mass $m_0^{}$, respectively. Given a true value of $m^{}_{\beta \beta}$ and an exposure ${\cal E}$, we can calculate the Bayes factor ${\cal B}$. Then, the contours of $\ln(\mathcal{B})$ in the plane of $m^{}_{\beta \beta}$ and ${\cal E}$ are plotted as thick solid (dashed) curves for the ``Low (High) NME" case. The black and orange curves stand for the contours of $\ln(\mathcal{B}) = 1$ and 2.5, respectively. In addition, there exists two blue curves for $\ln(\mathcal{B}) = -1$ in the upper part of the right panel. In the chosen ranges of $T^{0\nu}_{1/2}$ and ${\cal E}$, we have not found the contours of $\ln(\mathcal{B}) = 5$ and $-2.5$, which implies a limited power of discrimination between NO and IO even in a future experiment. Lastly, the red horizontal line is for ${m^{}_{\beta\beta}}|^{\rm IO}_{\rm min}$ that calculated by using the best-fit values of neutrino mixing parameters, while the dark (light) band is for the result by using $1\sigma$ ($3\sigma$) ranges.

After clarifying the conventions in Fig.~\ref{fg:NH_IH}, we are now at the position of drawing some physical implications from it. First of all, it is evident that the results depend on the prior distribution of the lightest neutrino mass $m_0^{}$. If the true $m^{}_{\beta\beta}$ is found to be large and well above the minimum ${m^{}_{\beta\beta}}|^{\rm IO}_{\rm min}$,  both IO and NO can explain such a large value, when a uniform prior on $m^{}_0$ is assumed. However, in the logarithmic prior case, the evidence will receive significant contributions from the region of much smaller $m^{}_0$, for which IO will result in a larger $m^{}_{\beta\beta}$ than NO. Therefore, IO is slightly preferred in this case, as indicated by the blue curves in the right panel.

However, when the true value of $m^{}_{\beta\beta}$ turns out to be close to ${m^{}_{\beta\beta}}|^{\rm IO}_{\rm min}$ or even smaller, the dependence on the priors of $m^{}_0$ becomes much weaker, and the results from both uniform and logarithmic priors tend to converge. In the following, we focus on this case, in which the results will not be much affected by the choices of priors. One can observe that both the contours of $\ln(\mathcal{B})= 1$ and $2.5$ indicate a preference for NO over IO. An immediate question is at which level we can exclude IO experimentally. In this connection, two comments are in order.

\begin{figure}
\begin{center}
\includegraphics[scale=0.75]{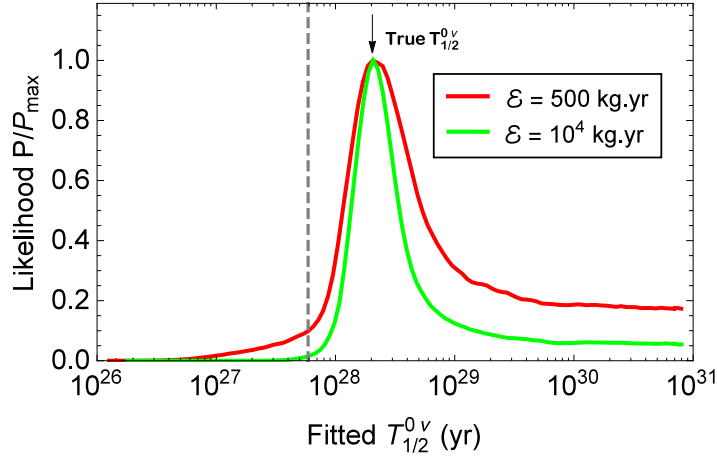}
\end{center}
\vspace{-0.5cm}
\caption{The likelihood function of the fitted $T^{0\nu}_{1/2}$ given the true value of $T^{0\nu}_{1/2}$ being $2 \times 10^{28}~{\rm yr}$. Red and green curves correspond to two different exposures $500~{\rm kg}\cdot{\rm yr}$ and $10^4~{\rm kg}\cdot{\rm yr}$, respectively, while the vertical dashed line indicates the upper bound of $T^{0\nu}_{1/2}$ in IO assuming the ``Low NME'' case. Background index is kept to be $10^{-4}$, and the other experimental parameters are the same as before.}
\label{fg:fittedT}
\end{figure}

First, as indicated by these contours, it is now evident that the exposure at least $500$ (or $2500$) kg~$\cdot$~yr needs to be accumulated in order to report a weak (or moderate) evidence for NO. According to the Jeffreys scale in Table~\ref{tb:Jeffreys}, the weak and strong evidence should be understood as a degree of belief of $75.0\%$ and $92.3\%$, respectively. This is due to the fact that when the exposure is too low, even though the true $m^{}_{\beta\beta}$ is much smaller than ${m^{}_{\beta\beta}}|^{\rm IO}_{\rm min}$, it cannot be precisely reconstructed from the data. As an example, we consider two different exposures, i.e., $\mathcal{E} = 500~{\rm kg}\cdot{\rm yr}$ and $10^4~{\rm kg}\cdot{\rm yr}$. In each case we first generate the data sample with the true value of $T_{1/2}^{0\nu}$ being $2 \times 10^{28}~{\rm yr}$, which corresponds to $m_{\beta\beta}^{} \sim 0.01~{\rm eV}$ below ${m^{}_{\beta\beta}}|^{\rm IO}_{\rm min}$ according to Eq.~(\ref{eq:rate}). We then fit these data samples with test values of $T_{1/2}^{0\nu}$, and show their likelihood functions (red and green curves) for both exposures in Fig.~\ref{fg:fittedT}, where we also draw the upper bound of $T_{1/2}^{0\nu}$ in IO (vertical dashed line) in the ``Low NME'' case. As one can see, in the $\mathcal{E}=500~{\rm kg}\cdot{\rm yr}$ case, although now the true value of $T_{1/2}^{0\nu}$ is above the IO upper bound, the likelihood function of the fitted $T_{1/2}^{0\nu}$ spreads out into the region allowed by IO. As a result, IO still has a non-negligible probability of explaining the data, weakening the preference of NO over IO. In contrast, when $\mathcal{E}=10^4~{\rm kg}\cdot{\rm yr}$, the probability of explaining the data with $T_{1/2}^{0\nu}$ in the region allowed by IO is much smaller, resulting in a higher degree of belief of preferring NO over IO. These findings agree with what we have observed by using a full statistical analysis in Fig.~\ref{fg:NH_IH}.
%%%%%%%%%%%%%%%%%%%%%%%%%%%%%%%%%%%%% Fig. 7 %%%%%%%%%%%%%%%%%%%%%%%%%%%%%
%%%%%%%%%%%%%%%%%%%%%%%%%%%%%%%%%%%%%%%%%%%%%%%%%%%%%%%%%%%%%%%%%%%%%%%%%

Second, in the future generation of $^{76}$Ge-based $0\nu\beta\beta$ experiments with an exposure of $10^4~{\rm kg}\cdot{\rm yr}$, we are able to reach a value of $\ln(\mathcal{B}) \gtrsim 2.5$, indicating a moderate evidence for NO. This conclusion is viable for both uniform and logarithmic priors. Fig.~\ref{fg:NH_IH} also confirms the fact that if the true $m^{}_{\beta\beta}$ is above ${m^{}_{\beta\beta}}|^{\rm IO}_{\rm min}$, then it is impossible to significantly distinguish between NO and IO within the $0\nu\beta\beta$ experiments.

\section{Summary and conclusion}

Current intensive experimental efforts on searching for $0\nu\beta\beta$ motivates us to perform an analysis of their physics prospects. In particular, we focus in the present work on the $^{76}$Ge-based $0\nu\beta\beta$ experiments, for its relatively simple background model. Three important aspects of this type of experiments are studied. First, we consider to set up the exclusion limits on the half life. It has been found that for the current generation of experiments (i.e., ${\cal E} = 400~{\rm kg}\cdot{\rm yr}$ and ${\rm BI} = 10^{-3}$) we can reach a half life of $4 \times 10^{26}$ yr, while $7 \times 10^{27}$ yr for the future one (i.e., ${\cal E} = 10^4~{\rm kg}$~$\cdot$~${\rm yr}$ and ${\rm BI} = 10^{-4}$). Second, we investigate the experimental ability of discovering a non-null signal. Our results indicate that if the true half life is below $\sim 10^{26}$ yr, one can report a strong evidence (or a degree of belief of $99.3\%$) for it with at least $50\%$ probability in the current generation of $^{76}$Ge-based experiments. For the future generation of experiments, the above limit can be increased to $\sim 2 \times 10^{27}$ yr. Lastly, we perform a statistical study on at which level the future generation of $^{76}$Ge-based experiments can discriminate two neutrino mass orderings. The primary results are given in Fig.~\ref{fg:NH_IH}. We find that for those experiments that have a very low background level of ${\rm BI} = 10^{-4}$, the exposure has to be at least $\sim 500$ ($\sim 2500$) kg~$\cdot$~yr before having the sensitivity to report a weak (moderate) evidence for NO against IO. According to the Jeffreys scale in Table~\ref{tb:Jeffreys}, the weak and moderate evidences should be understood as a degree of belief of $75.0\%$ and $92.3\%$, respectively. However, even in the future generation of experiments with an exposure of $10^4$ kg~$\cdot$~yr, a strong evidence for neutrino mass ordering cannot be obtained. In that case, we may achieve a definite conclusion on neutrino mass ordering by combining both $^{76}$Ge-based and other $0\nu\beta\beta$ experiments together. Nevertheless, the $0\nu\beta\beta$ experiments serve as another independent probe to neutrino mass ordering, which is complementary to the neutrino oscillation experiments~\cite{Qian:2015waa}.

Although we have concentrated on the $^{76}$Ge-based experiments, it should be emphasized that our analysis can directly be applied to other $0\nu\beta\beta$ experiments with distinct nuclear isotopes. Furthermore, a combined analysis of all types of experiments, as done in Refs.~\cite{Bergstrom:2012nx,Guzowski:2015saa}, will be helpful in extracting the valuable information not only on neutrino mass ordering but also on the absolute neutrino mass scale and Majorana CP-violating phases.

\section*{Acknowledgments}
\vspace{0.5cm}

This work was supported in part by the Innovation Program of the Institute of High Energy Physics under Grant No. Y4515570U1, by the
National Youth Thousand Talents Program, and by the CAS Center for
Excellence in Particle Physics (CCEPP).


\begin{thebibliography}{99}
%\cite{Furry:1939qr}
\bibitem{Furry:1939qr}
  W.~H.~Furry,
  %``On transition probabilities in double beta-disintegration,''
  Phys.\ Rev.\  {\bf 56}, 1184 (1939).
  %%CITATION = PHRVA,56,1184;%%
  %222 citations counted in INSPIRE as of 16 Jul 2015

%\cite{Barabash:2011mf}
\bibitem{Barabash:2011mf}
  A.~S.~Barabash,
  %``Double Beta Decay: Historical Review of 75 Years of Research,''
  Phys.\ Atom.\ Nucl.\  {\bf 74}, 603 (2011)
  [arXiv:1104.2714].
  %%CITATION = ARXIV:1104.2714;%%
  %24 citations counted in INSPIRE as of 16 juil. 2015

%\cite{Majorana:1937vz}
\bibitem{Majorana:1937vz}
  E.~Majorana,
  %``Theory of the Symmetry of Electrons and Positrons,''
  Nuovo Cim.\  {\bf 14}, 171 (1937).
  %%CITATION = NUCIA,14,171;%%
  %567 citations counted in INSPIRE as of 17 Aug 2015

%\cite{Agashe:2014kda}
\bibitem{Agashe:2014kda}
  K.~A.~Olive {\it et al.} [Particle Data Group Collaboration],
  %``Review of Particle Physics,''
  Chin.\ Phys.\ C {\bf 38}, 090001 (2014).
  %%CITATION = CHPHD,C38,090001;%%
  %1490 citations counted in INSPIRE as of 16 juil. 2015

%\cite{Bilenky:2014uka}
\bibitem{Bilenky:2014uka}
  S.~M.~Bilenky and C.~Giunti,
  %``Neutrinoless Double-Beta Decay: a Probe of Physics Beyond the Standard Model,''
  Int.\ J.\ Mod.\ Phys.\ A {\bf 30}, no. 04n05, 1530001 (2015)
  [arXiv:1411.4791].
  %%CITATION = ARXIV:1411.4791;%%
  %22 citations counted in INSPIRE as of 14 Aug 2015

%\cite{Pas:2015eia}
\bibitem{Pas:2015eia}
  H.~P\"{a}s and W.~Rodejohann,
  %``Neutrinoless Double Beta Decay,''
  arXiv:1507.00170.
  %%CITATION = ARXIV:1507.00170;%%
  %2 citations counted in INSPIRE as of 14 Aug 2015

%\cite{Kotila:2012zza}
\bibitem{Kotila:2012zza}
  J.~Kotila and F.~Iachello,
  %``Phase space factors for double-$\beta$ decay,''
  Phys.\ Rev.\ C {\bf 85}, 034316 (2012)
  [arXiv:1209.5722].
  %%CITATION = ARXIV:1209.5722;%%
  %110 citations counted in INSPIRE as of 14 Aug 2015

%\cite{Engel:2015wha}
\bibitem{Engel:2015wha}
  J.~Engel,
  %``Uncertainties in nuclear matrix elements for neutrinoless double-beta decay,''
  J.\ Phys.\ G {\bf 42}, no. 3, 034017 (2015).
  %%CITATION = JPAGA,G42,034017;%%
  %2 citations counted in INSPIRE as of 17 Aug 2015

%\cite{GomezCadenas:2011it}
\bibitem{GomezCadenas:2011it}
  J.~J.~Gomez-Cadenas, J.~Martin-Albo, M.~Mezzetto, F.~Monrabal and M.~Sorel,
  %``The Search for neutrinoless double beta decay,''
  Riv.\ Nuovo Cim.\  {\bf 35}, 29 (2012)
  [arXiv:1109.5515].
  %%CITATION = ARXIV:1109.5515;%%
  %113 citations counted in INSPIRE as of 14 Aug 2015

%\cite{Vergados:2012xy}
\bibitem{Vergados:2012xy}
  J.~D.~Vergados, H.~Ejiri and F.~Simkovic,
  %``Theory of Neutrinoless Double Beta Decay,''
  Rept.\ Prog.\ Phys.\  {\bf 75}, 106301 (2012)
  [arXiv:1205.0649].
  %%CITATION = ARXIV:1205.0649;%%
  %126 citations counted in INSPIRE as of 14 Aug 2015

%\cite{Maneschg:2015dja}
\bibitem{Maneschg:2015dja}
  W.~Maneschg,
  %``Review of neutrinoless double beta decay experiments: Present status and near future,''
  Nucl.\ Part.\ Phys.\ Proc.\  {\bf 260}, 188 (2015).

%\cite{GomezCadenas:2010gs}
\bibitem{GomezCadenas:2010gs}
  J.~J.~Gomez-Cadenas, J.~Martin-Albo, M.~Sorel, P.~Ferrario, F.~Monrabal, J.~Munoz-Vidal, P.~Novella and A.~Poves,
  %``Sense and sensitivity of double beta decay experiments,''
  JCAP {\bf 1106}, 007 (2011)
  [arXiv:1010.5112].
  %%CITATION = ARXIV:1010.5112;%%
  %43 citations counted in INSPIRE as of 14 Aug 2015

%\cite{Xing:2015zha}
\bibitem{Xing:2015zha}
  Z.~z.~Xing, Z.~h.~Zhao and Y.~L.~Zhou,
  %``How to interpret a discovery or null result of the $0\nu 2\beta$ decay,''
  Eur.\ Phys.\ J.\ C {\bf 75}, no. 9, 423 (2015)
  [arXiv:1504.05820].
  %%CITATION = ARXIV:1504.05820;%%
  %5 citations counted in INSPIRE as of 01 Oct 2015

%\cite{Abgrall:2013rze}
\bibitem{Abgrall:2013rze}
  N.~Abgrall {\it et al.} [Majorana Collaboration],
  %``The Majorana Demonstrator Neutrinoless Double-Beta Decay Experiment,''
  Adv.\ High Energy Phys.\  {\bf 2014}, 365432 (2014)
  [arXiv:1308.1633].
  %%CITATION = ARXIV:1308.1633;%%
  %31 citations counted in INSPIRE as of 14 Aug 2015


%\cite{Agostini:2013mzu}
\bibitem{Agostini:2013mzu}
  M.~Agostini {\it et al.} [GERDA Collaboration],
  %``Results on Neutrinoless Double-$\beta$ Decay of $^{76}$Ge from Phase I of the GERDA Experiment,''
  Phys.\ Rev.\ Lett.\  {\bf 111}, no. 12, 122503 (2013)
  [arXiv:1307.4720].
  %%CITATION = ARXIV:1307.4720;%%
  %199 citations counted in INSPIRE as of 16 juil. 2015

%\cite{Agostini:2015dna}
\bibitem{Agostini:2015dna}
  M.~Agostini, A.~Merle and K.~Zuber,
  %``Probing flavor models with Ge-76-based experiments on neutrinoless double-beta decay,''
  arXiv:1506.06133.
  %%CITATION = ARXIV:1506.06133;%%

\bibitem{LSGe}
  Neutrinoless Double Beta Decay Report to the Nuclear Science Advisory Committe, Apirl 24, 2014 ($\mathtt{http://science.energy.gov/np/nsac/reports}$).

%\cite{Caldwell:2006yj}
\bibitem{Caldwell:2006yj}
  A.~Caldwell and K.~Kroninger,
  %``Signal discovery in sparse spectra: A Bayesian analysis,''
  Phys.\ Rev.\ D {\bf 74}, 092003 (2006)
  [physics/0608249].
  %%CITATION = PHYSICS/0608249;%%
  %24 citations counted in INSPIRE as of 16 juil. 2015

%\cite{Cowan:2010js}
\bibitem{Cowan:2010js}
  G.~Cowan, K.~Cranmer, E.~Gross and O.~Vitells,
  %``Asymptotic formulae for likelihood-based tests of new physics,''
  Eur.\ Phys.\ J.\ C {\bf 71}, 1554 (2011)
  [Eur.\ Phys.\ J.\ C {\bf 73}, 2501 (2013)]
  [arXiv:1007.1727].
  %%CITATION = ARXIV:1007.1727;%%
  %748 citations counted in INSPIRE as of 18 juil. 2015


%\cite{Feroz:2008xx}
\bibitem{Feroz:2008xx}
  F.~Feroz, M.~P.~Hobson and M.~Bridges,
  %``MultiNest: an efficient and robust Bayesian inference tool for cosmology and particle physics,''
  Mon.\ Not.\ Roy.\ Astron.\ Soc.\  {\bf 398}, 1601 (2009)
  [arXiv:0809.3437].
  %%CITATION = ARXIV:0809.3437;%%
  %395 citations counted in INSPIRE as of 14 Aug 2015

%\cite{Feroz:2007kg}
\bibitem{Feroz:2007kg}
  F.~Feroz and M.~P.~Hobson,
  %``Multimodal nested sampling: an efficient and robust alternative to MCMC methods for astronomical data analysis,''
  Mon.\ Not.\ Roy.\ Astron.\ Soc.\  {\bf 384}, 449 (2008)
  [arXiv:0704.3704].
  %%CITATION = ARXIV:0704.3704;%%
  %291 citations counted in INSPIRE as of 14 Aug 2015

%\cite{Feroz:2013hea}
\bibitem{Feroz:2013hea}
  F.~Feroz, M.~P.~Hobson, E.~Cameron and A.~N.~Pettitt,
  %``Importance Nested Sampling and the MultiNest Algorithm,''
  arXiv:1306.2144.
  %%CITATION = ARXIV:1306.2144;%%
  %67 citations counted in INSPIRE as of 14 Aug 2015

\bibitem{SSbook}
  D.~S.~Sivia and J.~Skilling, \emph{Data Analysis: a Bayesian Tutorial} (Oxford University Press, 2006), 2nd ed.

%\cite{Cowan:1998ji}
\bibitem{Cowan:1998ji}
  G.~Cowan,
  \emph{Statistical data analysis},
  Oxford, UK: Clarendon (1998) 197 p
  %36 citations counted in INSPIRE as of 14 Aug 2015

\bibitem{KR}
  R.~E.~Kass and A.~E.~Raftery, Journal of the American Statistical Association {\bf 90}, 791 (1995).

\bibitem{Jeffreys}
  H.~Jeffreys, \emph{The Theory of Probability} (Oxford University Press, 1961), 3rd ed.

\bibitem{Jeffreys_mod}
  M.~Hobson \emph{et al.}, eds., \emph{Bayesian methods in cosmology} (Cambridge University Press, 2010).

%\cite{Trotta:2008qt}
\bibitem{Trotta:2008qt}
  R.~Trotta,
  %``Bayes in the sky: Bayesian inference and model selection in cosmology,''
  Contemp.\ Phys.\  {\bf 49}, 71 (2008)
  [arXiv:0803.4089 [astro-ph]].
  %%CITATION = ARXIV:0803.4089;%%
  %172 citations counted in INSPIRE as of 12 Nov 2015

%\cite{Gonzalez-Garcia:2014bfa}
\bibitem{Gonzalez-Garcia:2014bfa}
  M.~C.~Gonzalez-Garcia, M.~Maltoni and T.~Schwetz,
  %``Updated fit to three neutrino mixing: status of leptonic CP violation,''
  JHEP {\bf 1411}, 052 (2014)
  [arXiv:1409.5439].
  %%CITATION = ARXIV:1409.5439;%%
  %146 citations counted in INSPIRE as of 14 Aug 2015

%\cite{Guzowski:2015saa}
\bibitem{Guzowski:2015saa}
  P.~Guzowski, L.~Barnes, J.~Evans, G.~Karagiorgi, N.~McCabe and S.~Soldner-Rembold,
  %``Combined limit on the neutrino mass from neutrinoless double-¦Â decay and constraints on sterile Majorana neutrinos,''
  Phys.\ Rev.\ D {\bf 92}, no. 1, 012002 (2015)
  [arXiv:1504.03600].
  %%CITATION = ARXIV:1504.03600;%%
  %3 citations counted in INSPIRE as of 20 ao?t 2015

%\cite{Qian:2015waa}
\bibitem{Qian:2015waa}
  X.~Qian and P.~Vogel,
  %``Neutrino Mass Hierarchy,''
  Prog.\ Part.\ Nucl.\ Phys.\  {\bf 83}, 1 (2015)
  [arXiv:1505.01891].
  %%CITATION = ARXIV:1505.01891;%%
  %3 citations counted in INSPIRE as of 14 Aug 2015

%\cite{Bergstrom:2012nx}
\bibitem{Bergstrom:2012nx}
  J.~Bergstr\"{o}m,
  %``Combining and comparing neutrinoless double beta decay experiments using different nuclei,''
  JHEP {\bf 1302}, 093 (2013)
  [arXiv:1212.4484].
  %%CITATION = ARXIV:1212.4484;%%
  %11 citations counted in INSPIRE as of 21 ao?t 2015

\end{thebibliography}
\end{document}